\documentclass[preprint,showpacs,aps]{revtex4}
\newcommand{\dslash}{d\!\!{}^{-}}
\begin{document}
\tolerance=5000
\def\be{\begin{equation}}
\def\ee{\end{equation}}
\def\bea{\begin{eqnarray}}
\def\eea{\end{eqnarray}}

\title{Entropy and anisotropy}

\author{Francisco J. Hern\'andez and Hernando Quevedo}
\email{fmoreno, quevedo@nucleares.unam.mx} 
\affiliation{   Instituto
de Ciencias Nucleares,
     Universidad Nacional Aut\'onoma de M\'exico \\
     P.O. Box 70-543, 04510 M\'exico D.F., M\'exico}

\date{\today}% It is always \today, today,

\begin{abstract}
We address the problem of defining the concept of entropy for 
anisotropic cosmological models. In particular, we analyze 
for the Bianchi I and V models the entropy which follows from 
postulating the validity of the laws of standard thermodynamics
in cosmology. Moreover, we analyze the Cardy-Verlinde construction 
of entropy and show that it cannot be associated with the one following 
from relativistic thermodynamics. 
\end{abstract}

\maketitle

\section{Introduction}

Entropy is a very important concept in physics. As a thermodynamic variable
it should be present in any physical system to which energy, in any of its
forms, can be associated. The origin of thermodynamics is purely 
phenomenological and, consequently, is based on certain laws which 
are the result of  experiments usually performed on simple systems.
It is therefore a difficult task to generalize these laws to include
cases in which experiments are not available. For instance, it is not
completely clear how to handle relativistic systems from the point of
view of thermodynamics. This is especially difficult in the case of 
gravitational systems in which the concept of energy is not well-defined.
Nevertheless, under certain assumptions gravitational fields can be 
treated as thermodynamical systems, i.e. systems in which the standard 
laws of thermodynamics are supposed to be valid. For a review of the 
main aspects of relativistic thermodynamics see, for instance, \cite{mtw}.
In the case of a simple Friedman-Robertson-Walker (FRW) universe,
relativistic thermodynamics gives certain values for the 
thermodynamic variables which are in agreement with  
physical expectations \cite{kolb}. 
 
Nevertheless, it 
is quite possible to find cosmological configurations in which mathematically 
well-defined state and thermodynamic variables predict unphysical behaviors.
Indeed, it has been shown \cite{kqs} that the thermodynamic variables 
obtained by applying the laws of relativistic thermodynamics 
to inhomogeneous cosmological models
lead to unphysical temperature evolution laws, and that physical relevant
behaviors can be expected only when all the inhomogeneities vanish, i.e.,
in the FRW limit. This result obviously points out to an inconsistency between
cosmology in general relativity and the first law of thermodynamics, 
at least when inhomogeneities are present. The natural question arises 
whether this inconsistency persists in the case of homogeneous (anisotropic)
cosmological models. In this work we will show that cosmological 
anisotropies can be treated in a consistent manner in the context
of relativistic thermodynamics.

Recently, Verlinde \cite{ver} proposed an alternative approach to the 
concept of entropy, based on a formal analogy between the field equations
for a FRW cosmology and the thermodynamic formulas of conformal field 
theory (CFT). This analogy seems to be related to the holographic
principle according to which for a given volume one can associate 
a maximal amount of entropy which corresponds to the entropy of 
the largest black hole that can be fitted inside this volume  
\cite{thooft,suss}. Furthermore, the entropy of black holes
has been calculated recently \cite{youm99,ash98} by counting
microscopic states.  Although these computations start from very different
physical concepts,  they make use of techniques known in two-dimensional
CFT. At the first sight this seems to be only a useful trick --the Cardy
entropy formula \cite{Cardy} allows to easily count  states in  two-dimensional
CFT-- but some recent results \cite{malstr97,lars97} suggest that CFT's 
offer a model-independent description of black hole thermodynamics at low
energies. The universality of this description might be related to 
the simple common feature that the algebra of diffeomorphisms at the black
hole horizon has a conformal structure \cite{carlip99}. In light of
these results it seems reasonable to expect that the Cardy entropy formula
could have some applications in cosmology as proposed by Verlinde \cite{ver}
for FRW models. The Cardy-Verlinde procedure for defining entropy has been generalized to 
anisotropic Bianchi IX cosmologies in \cite{opq03}.

In this paper, we analyze the entropy of homogeneous  
cosmological models with a perfect fluid as source. 
We first review in Section \ref{sec:class}
an alternative approach to classical 
thermodynamics which is based upon the contact structure of
the thermodynamic phase space. This geometric approach 
allows a simple generalization to the case of relativistic 
systems, and makes it particularly easy to interpret relativistic 
thermodynamics. We find the sufficient and necessary conditions 
which need to be fulfilled in order 
to determine entropy and temperature for a given cosmological 
model, and show in Section \ref{sec:hom}
that in the case of homogeneous configurations they
are identically satisfied. This result is used to 
compute explicit expressions for the entropy and temperature
of homogeneous models and we show that they correspond to 
an adiabatic process, and evolve
in a physically reasonable manner
 when compared with their  FRW counterparts.
Then in Section \ref{sec:cardy} we  find for Bianchi I and V models
the corresponding 
CFT entropy by using the Cardy-Verlinde formula. It is shown
that this entropy does not satisfy the adiabatic property which
follows from energy conservation and relativistic thermodynamics.
We conclude in Section \ref{sec:con} 
that this result is not enough to dismiss Cardy's
entropy as unphysical.

\section{Classical thermodynamics of equilibrium states}
\label{sec:class}

In this section we find the conditions which must be satisfied in order 
to define the entropy for a given thermodynamic system. For
the sake of simplicity, we limit ourselves here to the case of a 
monocomponent simple system. According to Hermann's geometric approach 
\cite{hermann}, one usually begins with the introduction of 
a suitable {\it thermodynamic phase space} ${\cal T}$ which in this 
case is a 5-dimensional manifold, topologically equivalent to ${\cal R}^5$. 
In ${\cal T}$ we can introduce coordinates $\{U,T,S,P,V\}$ which correspond
to the thermodynamic variables of internal energy, temperature, entropy, pressure,
and volume, respectively. If we demand smoothness, 
at each point $x$ of ${\cal T}$ we can construct 
the tangent $T_x{\cal T}$ and cotangent $T_x^*{\cal T}$ manifolds in the standard
manner so that vectors, tensors, and differential forms are well-defined 
geometric objects. In particular, we introduce the fundamental Gibss 1-form 
\be
\Theta = dU - T dS + P dV \ ,
\label{gibbs}
\ee
where $d$ represents the operator of exterior derivative. This is a very general 
construction in which all simple thermodynamic systems can be represented. 
To differentiate one thermodynamic system from another, one usually specifies an
equation of state which is a relationship between different thermodynamic 
variables. Alternatively, one can specify the fundamental equation from which
all the equations of state can be derived \cite{callen}. In the energy representation 
we are using for the thermodynamic phase space, the fundamental equation 
relates the internal energy $U$ with the state thermodynamic variables. In principle 
one can take any pair of variables $\{T,S,P,V\}$ as state variables. The only 
condition is that they must be well-defined in the corresponding submanifold of 
${\cal T}$. For later use we choose a fundamental equation of the form $U=U(P,V)$.

Although the following definition makes use of a specific fundamental 
equation, it can be shown that it does not depend on it \cite{burke}. 
A simple thermodynamic equilibrium system corresponds to a two-dimensional 
submanifold ${\cal E} \subset {\cal T}$ defined by the smooth mapping 
$ \varphi : \  {\cal E} \  \rightarrow  {\cal T}$ with 
\be
\varphi :  (P,V) \longmapsto [U(P,V), T(P,V), S(P,V), P, V] \ .
\label{map}
\ee
such that the pull-back $\varphi^*$ of the Gibbs 1-form vanishes, i.e.
\be
\varphi^*(\Theta) = 0 \ ,
\label{pullback}
\ee
and the convexity condition is satisfied:
\be
\frac{\partial^2 U }{\partial X^A X^B} \geq 0 \ ,
\ee
where $X^A=(P,V)$. In the energy representation we are using here, 
%%%%%
the convexity 
condition leads to 
the {\it second law of thermodynamics} and  reduces to its standard form in the entropy representation.
%%%%%%%

Using Eq.(\ref{gibbs}), condition (\ref{pullback}) reads  
\be
\frac{\partial U}{\partial P} = T \frac{\partial S}{\partial P} \ , \quad
\frac{\partial U}{\partial V} = T \frac{\partial S}{\partial V} - P \ .
\ee 
Consequently, on the space of equilibrium states of a given thermodynamic 
system we obtain the {\it first law of thermodynamics} %%%%
\footnote{In fact, this is Gibbs' equation which in cosmology  
is mistakenly referred to as the first law.} 
\be
TdS = d U + P dV \ ,
\label{gibbsdown}
\ee
with $U=U(P,V)$. This construction shows that if a thermodynamic
system is considered by means of its fundamental equation $U=U(P,V)$ 
as a submanifold 
${\cal E}$ of the general thermodynamic phase space ${\cal T}$, then
the variables $T$ and $S$ on ${\cal E}$ are determined through the differential relationship 
(\ref{gibbsdown}). The question arises whether in general it is possible 
to integrate the first law of thermodynamics as derived in (\ref{gibbsdown}).
The answer can easily be found by using 
Frobenius' theorem according to which for the differential 1-form 
\be
\Omega := dU + PdV
\ee
to be integrable it is necessary and sufficient that \cite{tresdamas}
\be
\Omega \wedge d \Omega = 0 \ ,
\label{intcond}
\ee
where the wedge represents the exterior product. On the 2-dimensional manifold
${\cal E}$ this condition is trivially satisfied since any 3-form on ${\cal E}$
vanishes identically. Consequently, it must be always possible to find functions
$T=T(P,V)$ and $S=S(P,V)$ such that (\ref{gibbsdown}) is satisfied. This is in 
accordance with the definition of the embedding mapping $\varphi$ as given in 
(\ref{map}). 
Notice that the sufficient condition $d\Omega =0$ is not
satisfied in general; if it were satisfied, we could find a function, say $Q=Q(P,V)$,
such that $dQ=\Omega=dU+PdV$, an expression which is obviously 
not true. This relationship 
is often written as $\dslash Q = dU+PdV$ to emphasize that the right-hand side
is not an exact 1-form. 

Another important element of thermodynamics is the Euler identity which is 
a consequence of the existence of extensive thermodynamic variables which 
can be used as coordinates in the space of equilibrium states. In the case of
the simple system we are studying here the extensive variables are entropy and
volume. Consider a mapping $\tilde\varphi: \cal{E} \rightarrow \cal{T}$ with 
\be
\tilde \varphi : (S,V) \longmapsto [U(S,V), T(S,V), S, P(S,V), V] \ ,
\label{map1}
\ee
so that in these variables the condition $\tilde\varphi^*(\Theta) =0$ yields 
\be
\frac{\partial U}{\partial S} = T\ ,\quad 
\frac{\partial U}{\partial V} = -P \  .
\label{equi1}
\ee
The thermodynamical potential $U=U(S,V)$ satisfies the homogeneity condition
$U(\lambda S, \lambda V) = \lambda^\beta U(S,V)$ for constants $\lambda$ and 
$\beta$. The variables $S$, $V$, and $U$ are called extensive, sub-extensive or 
supra-extensive  if $\beta=1$, $\beta <1$ or $\beta >1$, respectively. From the
homogeneity condition we obtain
\be
\frac{\partial U(\lambda S, \lambda V)}{\partial (\lambda S)} 
\frac{\partial (\lambda S)}{\partial \lambda} 
+
\frac{\partial U(\lambda S, \lambda V)}{\partial (\lambda V)}
\frac{\partial (\lambda V)}{\partial \lambda} 
 =
\beta \lambda^{\beta -1} U(S,V) \ .
\ee
Putting $\lambda = 1$ and using the relations (\ref{equi1}), from the last equation 
we get the Euler identity
\be
\beta U - TS + PV =0 \ .
\label{euler}
\ee
Furthermore, calculating the exterior derivative of the Euler identity and using
the first law (\ref{gibbsdown}), we obtain the Gibbs-Duhem relation
\be
S d T - V dP + (1-\beta) d U = 0 \ .
\label{gibduh}
\ee

The above geometric approach to thermodynamics is based only on the embedding 
structure of the thermodynamic phase space and the space of equilibrium states.
A more general structure can be obtained by introducing Riemannian metrics
on both spaces and comparing them by imposing invariance with respect to
Legendre transformations. The resulting geometrothermodynamical 
approach allows to handle certain 
aspects of thermodynamic systems in terms of geometric objects \cite{hq06}.

%%%%%%%%%%%%%%%%%%%%%%%%%%%%%%%%%%%%%%%%%%%%%%%%%%%%%%%%%%%%%%%%
%%%%%%%%%%%%%%%%%%%%%%%%%%%%%%%%%%%%%%%%%%%%%%%%%%%%%%%%%%%%%%% 
\section{Relativistic thermodynamics}
\label{sec:rel}

The generalization of thermodynamics to gravitational systems is a delicate
procedure which must take into account 
the invariance of certain thermodynamic variables
with respect to measurements carried out by different observers 
(see, for  instance, \cite{mtw}, for a lucid introductory review).
In its final form, relativistic thermodynamics consists in 
imposing the fulfillment in curved spacetimes
of the first law of thermodynamics (\ref{gibbsdown}), 
the Euler identity (\ref{euler}) and the Gibbs-Duhem
relation (\ref{gibduh}), 
whereas the second law is imposed 
in the form of an entropy current \cite{mtw}
(see below).  

In the geometric language introduced in the last section
the passage to the relativistic generalization is quite simple.
The first law of thermodynamics in a curved spacetime 
corresponds to assuming the validity of (\ref{gibbsdown}) 
with the exterior derivative operator acting on functions
which depend on the spacetime coordinates, i.e. we assume
that $P=P(x^\mu)$ and $V=V(x^\mu)$ with $\mu=0,1,2,3$. 
Then the {\it first law of relativistic thermodynamics} reads
\be
T S_{,\mu}\, dx^\mu = (U_{,\mu} + P V_{,\mu})\,dx^\mu \ ,
\label{gibbsdown1}
\ee
where the comma represents partial derivatives. 
 The first thing we notice
now is that the integrability condition (\ref{intcond}) is no
longer identically satisfied since it represents now a 3-form
on a 4-dimensional manifold. In fact, if $\Omega_\mu$ represents
the components of the 1-form $\Omega$ in the coordinate basis
$\{dx^\mu\}$, the integrability condition is equivalent to
\be
\Omega_{[\mu}\Omega_{\nu,\tau]} = 0 \ ,
\label{ic1}
\ee
where the square brackets
indicate antisymmetrization.  If this condition is not satisfied,
there are no functions $T(x^\mu)$ and $S=S(x^\mu)$ such that the
first law of thermodynamics (\ref{gibbsdown}) is fulfilled. Consequently, 
in relativity theory one could in principle find gravitational
systems which cannot be treated as thermodynamic systems. 
In some sense, this is not surprising since this approach 
implies that the corresponding system must be in equilibrium as 
a thermodynamic system, and it is not difficult to imagine 
gravitational systems with no equilibrium states at all.
%%%%%%%%%%%%%%
On the other hand, the question arises whether there are 
gravitational systems for which the integrability condition 
is satisfied. If the answer is positive, Frobenius' theorem 
guarantees the existence of functions $T(x^\mu)$ and $S(x^\mu)$ 
which fulfill Eq.(\ref{gibbsdown1}); then, these functions
can be considered as physically meaningful, i.e., as
temperature and entropy of the system, 
if they satisfy the remaining thermodynamic laws.
%%%%%%%%%%  
This is exactly the
question that was analyzed in a series of works \cite{kqs}
with the result that there exists gravitational configurations
where mathematically well-defined thermodynamic variables
predict unphysical behaviors. 

In this work we will focus on gravitational systems corresponding
to cosmological models with a perfect fluid source 
\be
T_{\mu\nu} = (\rho+p)u_\mu u_\nu + p g_{\mu\nu} \ ,
\ee
where $p$, $\rho$ and $u_\mu$ are the pressure, energy density 
and 4-velocity, respectively. The conservation law for this
energy-momentum tensor can be written as
\be
\dot \rho + (\rho+p) \Theta = 0 \ , \quad
h^\nu_\mu p_{,\nu} + (\rho + p) \dot u _\mu = 0  \ ,
\label{claw}
\ee
where $\dot \rho = u^\mu \rho_{,\mu}$, 
$\Theta = u^\mu_{;\mu}$ is the expansion, $\dot u ^\mu = u_{\nu;\mu} u ^\nu$
is the 4-acceleration and $h^\mu_\nu = \delta^\mu_\nu + u_\nu u^\mu$ is the
projection tensor. The internal energy is $U=\rho V$ and 
\be
\Omega = [ V \rho_{,\mu} + (\rho+p) V_{,\mu} ] dx^\mu \ .
\label{flaw2}
\ee
The further assumption that $\Omega = T S_{,\mu} dx^\mu$ 
corresponds to the first law of thermodynamics for the perfect fluid. 
A straightforward calculation shows that the integrability condition 
(\ref{ic1}) implies that
\be
\rho_{[,t}p_{,i}V_{,j]} = 0 \ , \quad
\rho_{[,i}p_{,j}V_{,k]} = 0 \ ,
\label{ic2}
\ee
where $t$ is the time-coordinate and small latin
indices denote spatial coordinates. Since the 
thermodynamic variables $\rho$ and $p$ are related through Einstein's
equations, it is clear that Eqs.(\ref{ic2})  are not necessarily identically 
satisfied. Notice that the fundamental equation $U=U(P,V)$ in the
case of the perfect fluid under consideration reduces 
to $\rho=\rho(p)$, i.e. to a barotropic equation of state.

The second law  is postulated for the entropy current $(s u^\mu)$ 
with $s= S/V$ in the form 
\be
(su^\mu)_{;\mu} \geq 0
\ee
where the equality holds in the case of no entropy production. 
Usually, the
second law is considered together with the condition of conservation
of  the particle number density $n=1/V$:
\be
\left({nu^\mu}\right)_{;\mu} =0  \ .
\ee

%%%%%%%%%%%%%%%%%%%%%%%%%%%%%%%%%%%%%%%%%%%%%%%%%%%%%%%%%%% 
%%%%%%%%%%%%%%%%%%%%%%%%%%%%%%%%%%%%%%%%%%%%%%%%%%%%%%%%%%
\section{Homogeneous cosmological models}
\label{sec:hom}

Let us consider an non-rotational, homogeneous perfect fluid. The 4-velocity
can be shown to be hypersurface orthogonal and therefore there exist 
local comoving coordinates $(t,x^i)$ such that 
\be
ds^2 = N^2 dt^2 - g_{ij} dx^idx^j\ , \quad 
u^\mu = N^{-1}\delta^\mu_t\ ,\quad 
\dot u _i = (\ln N)_{,\mu}\delta^\mu_i \ ,\quad
h_{\mu\nu} = g_{ij}\delta^i_\mu \delta^j_\nu \ .
\label{lel}
\ee
Considering that all metric coefficients are independent of the spatial 
coordinates, the conservation law (\ref{claw}) reduces to 
\be
\rho_{,t} + (\rho +p) (\ln\sqrt{\Delta})_{,t} = 0 \ ,\quad
p_{,i} = 0 \ ,
\label{claw1}
\ee
where $\Delta = \det(g_{ij})$. It is possible to introduce a new time 
coordinate so that the lapse function $N=1$. We choose such a time
coordinate and denote by a dot the derivative with respect to it.
For simplicity we denote this new time coordinate again as $t$.  
Moreover, we will limit ourselves to perfect fluids that satisfy 
a barotropic equation of state, i.e. $p=\omega \rho$, where $\omega$
is a constant. Then, Eq.(\ref{claw1}) can be integrated and yields
\be
p=p(t)\ ,\quad
\rho=\rho_0 \Delta^{-(1+\omega)/2} \ ,
\ee
where $\rho_0$ is a positive constant. 

We now analyze the thermodynamic variables. According to Eq.(\ref{ic2}),
the integrability conditions to determine temperature and entropy are identically
satisfied for time-dependent functions $p$ and $\rho$. This means that 
there must exist mathematical expressions for $T$ and $S$ satisfying 
Eq.(\ref{flaw2}), i.e. 
\be
T \dot S = V\left[\dot \rho + (p+\rho)\frac{\dot V}{V }  \right] \ .
\label{flwa3}
\ee
The physical volume in the case of the metric (\ref{lel}) can be defined
as $V=\int\sqrt{{\rm det}(g_{ij})}d^3 x =\kappa \sqrt{\Delta}$ where $\kappa$ is a constant
which can be chosen as $\kappa=1$, without loss of generality.
It follows then from (\ref{flwa3}) and (\ref{claw1}) that the expansion
in this cosmological model corresponds to an adiabatic process, i.e.
\be
\dot S = 0 \ .
\label{adia}
\ee

On the other hand, from the Euler identity (\ref{euler}) with $U=\rho V$ and $\beta=1$
we have that
\be 
S = \frac{p+\rho}{T} V = \frac{(1+\omega)\rho}{T} \Delta^{1/2} = \frac{(1+\omega)\rho_0}{T\Delta^{\omega/2}}
\ee
for a barotropic state equation. Consequently, the adiabaticity condition 
(\ref{adia}) implies  that 
\be
S = \frac{(1+\omega)\rho_0}{T_0} \ , \quad
T = T_0 \Delta^{-\omega/2}\ ,
\label{tvar}
\ee
where $T_0$ is a positive constant. The physical relevance of these 
expressions can be derived by comparison with the corresponding expressions 
in FRW cosmologies. We see that the entropy for anisotropic models is a constant 
that always can  be made to coincide with the corresponding FRW value. The
temperature evolves as the physical volume $\Delta$ in the same way as in the 
FRW case. Consequently, the behavior of the anisotropic temperature will 
coincide with the limiting FRW case if the physical volume behaves similarly.
This can be shown to be true in general homogeneous models. 
In particular, for the Bianchi IX model it was shown in \cite{opq03} that
the different anisotropies essentially do not affect the dynamical behavior
of the physical volume which turns out to be determined only by the different
scale factors. This shows that the mathematical expressions for entropy and 
temperature (\ref{tvar}), which follow from the laws of relativistic thermodynamics,
are physically meaningful. 
  
\section{The Cardy entropy}
\label{sec:cardy}

In two-dimensional CFT, the Cardy formula allows to count microscopic 
states in a particularly
easy manner and leads to an explicit value for the entropy \cite{Cardy}
\be
 S_C = 2 \pi \sqrt{ {c\over 6} \left(L_0 -{c\over 24}\right)} \ ,
\label{cardy}
\ee
where $c$ is the central charge and $L_0$ the eigenvalue of the 
Virasoro operator. Verlinde \cite{ver} postulated the universal
validity of this formula and found a surprising link with 
 Friedman's equations. In \cite{opq03} it was shown that the Cardy formula
 can be generalized to the case of closed anisotropic cosmologies  
described by the Bianchi IX models. To see if this generalization 
can be extended to include plane and open anisotropic cosmologies, 
let us consider the Bianchi I and V models whose metric can be
written as
\begin{equation}\label{mbianchi}
ds^{2}= dt^{2}-a_{1}(t)^{2}dx^{2}-
a_{2}(t)^{2}e^{2\alpha x}dy^{2}
-a_{3}(t)^{2}e^{2\alpha x}dz^{2}.
\end{equation}
The metric for the Bianchi I geometry formally corresponds to the case 
$\alpha = 0$, while for the Bianchi  V case we have $\alpha = 1$.
To study the correspondence between  Cardy's formula and these cosmological models 
we only need the Hamiltonian constraint which can be expressed as 
\be\label{constham}
H_{1}H_{2}+H_{1}H_{3}+H_{2}H_{3}+ F(a_1,a_2,a_3) = 8 \pi G  \rho
\ee
with the directional Hubble parameters defined as $H_i = \dot a _ i/a_i$. 
Here $G$ is Newton's gravitational constant and $\rho$ is the energy density
of the perfect fluid. 
For convenience we introduced the function $F(a_1,a_2,a_3)$ which takes different
values for the different Bianchi models we are considering here (see Table I). 
It is also convenient to introduce a constant factor $k$ into the Cardy formula 
in order to include all different cases. Then
\be
 S_C = 2 \pi \sqrt{ {c\over 6} \left(L_0 -k{c\over 24}\right)} \ ,
\label{cardyk}
\ee
where $k=0,-1,1$ for Bianchi I, V and IX  models, respectively. 
%%%%%%%%%%%%%
An equivalent relationship was proposed by Youm \cite{youm02} in an attempt
to generalize Cardy formula to include the different types
of FRW cosmologies.  Clearly, in the isotropic limit of the Bianchi I, V and
IX models we recover the corresponding spatially flat, open an closed 
FRW universes analyzed in \cite{youm02}. 
%%%%%%%%%%%%%%%%%%%

Now, the idea is to identify $S_C$, $L_0$, and $c$ such that the Cardy 
formula (\ref{cardyk}) reduces formally to the Hamiltonian constraint
(\ref{constham}). It is easy to see that the identification is unique 
and corresponds to 
\be
S_{C} = \frac{1}{2 \sqrt{3} G}
\sqrt{H_{1}H_{2}+H_{1}H_{3}+H_{2}H_{3}}\;V \ ,
\label{cardyanis}
\ee
\be 
L_0=\frac{1}{3}\tilde{a} E \, \quad
c = \frac{3}{\pi G }\,\frac{V}{\tilde{a}} \ ,
\label{l0c}
\ee
where $V$ is the physical volume, $V=a_1a_2a_3$, 
the total energy is $E=\rho V$,
 and $\tilde a$ is a function of the directional scale factors.
We conclude that Cardy's formula can be generalized 
to include the cases of plane and open cosmologies. The explicit
form of Cardy's entropy remains the same for all cases considered
and differences appear only at the level of the central charge and
Virasoro operator.  
In Table I we present the explicit values of these quantities for all
cases investigated. We see that Cardy's formula postulates an explicit value for the CFT's entropy of 
homogeneous cosmologies. This has been shown explicitly 
only for Bianchi I, V, and  IX models,
but the generality of our results seems to indicate that they 
are valid for any Bianchi cosmology. 

%%%%%%%%%%%%%%%
The modifications  due to the presence of anisotropies can be 
derived from Eqs.(\ref{cardyanis}) and (\ref{l0c}).
In fact, these relationships can be obtained by introducing an 
effective Hubble parameter 
$H\rightarrow \tilde H = \sqrt{(H_{1}H_{2}+H_{1}H_{3}+H_{2}H_{3})/3}$
and an effective scale factor $a\rightarrow \tilde a$ (see Table I) 
in the original FRW values, which correspond to the limiting 
case $a_1 = a_2 = a_3 = a$. We conclude that in the case of anisotropic 
cosmologies the identification of the Cardy entropy with the 
Hamiltonian constraint can be performed just by introducing effective 
anisotropic parameters. 
%%%%%%%%%%%%%%%%%%%%%%%%%%%%

\begin{table}[htdp]
%\caption{CFT's entropy for Bianchi cosmologies}
\begin{center}
\begin{tabular}{|c|c|c|c|c|}
\hline
Bianchi &  CFT's entropy  & Constraint & Parameters'  & identification   \\
%\hline
%\hline 
Type &  $ S_C = 2 \pi \sqrt{ {c\over 6} \left(L_0 -k{c\over 24}\right)}$ 
&  $F(a_1,a_2,a_3)$ & $ L_0 $ & $c$ \\
\hline
 I &  $ 2\pi \sqrt{\frac{c}{6}\,L_{0}}$ & 0 
 & $\frac{1}{3}{a_1} E $&  $\frac{3}{\pi G }\frac{V}{a_1}$ \\
& &&&\\
 V& $ 2 \pi \sqrt{ {c\over 6} \left(L_0 +{c\over 24}\right)}$
& -$\frac{3}{a_{1}^{2}}$
& $\frac{1}{3}{a_1} E $&  $\frac{3}{\pi G }\frac{V}{a_1}$  \\
& &&&\\
 IX  &
 $ 2\pi \sqrt{ {c\over 6} \left(L_0 -{c\over 24}\right)}$&
 $\frac{3}{a_{1}^{2}}\left(1+\frac{1}{3}\epsilon^2 \right)\;\;$
 & $ \frac{a_{1}E}{3\sqrt{1+\epsilon^{2}/3}}$  
 &
 $\frac{3 \sqrt{1+\epsilon^{2}/3}\ V }{\pi G a_1}$
 \\
& &&&\\
\hline
\end{tabular}
\end{center}
\caption{{\bf CFT's entropy for Bianchi cosmologies}. 
In all cases Cardy's formula gives a definite value for the entropy $S_C$ as given in 
Eq.(\ref{cardyanis}), while the values of parameters $L_0$ and $c$ are used to reproduce
the Hamiltonian constraint. 
In the case of Bianchi IX models the explicit form of $\epsilon = \epsilon(a_1,a_2,
a_3)$ is given in \cite{opq03}. 
%%%%%%%%%%%%%%%
Notice that the chosen value of $\tilde a$ seems 
to single out the scale factor $a_1$, i.e., the anisotropy in the $x-$direction. 
This is only a matter of convention because similar expressions can be written for
$a_2$ and $a_3$, with the corresponding changes in the expressions for $\epsilon$
and $\tilde a$. Therefore, any direction could have been chosen for writing down
the explicit expressions.
%%%%%%%%%%%%%%%%  
}
\label{tab1}
\end{table}%

We now turn back to the study of the compatibility of Cardy's entropy with the
laws of relativistic thermodynamics. If we try to identify Cardy's entropy $S_C$
with the entropy $S=$const we obtained from the thermodynamic approach of Section 
\ref{sec:hom}, the first thing we can notice is that
Cardy's entropy does not satisfy the adiabaticity condition $\dot S _C=0$. 
At the first sight, this could be a sufficient reason for considering  
Cardy's cosmological entropy as unphysical since it is not compatible with 
the laws of relativistic thermodynamics. However, we believe that it is necessary
to perform a deeper analysis, before trying to make definite conclusions. 

On the one hand, one can try to take into account other properties of
CFT's entropy to see if it is possible to arrive to
an adiabatic expression. As pointed out by Verlinde \cite{ver}, Cardy's entropy is 
characterized by its sub-extensive nature. Euler's identity (\ref{euler}) 
for CFT's entropy $S_C$ turns out to be valid only for $\beta = 1/3$, i.e. 
\be 
\frac{1}{3}U_C - T S_C + PV =0 \ ,
\ee
where $U_C$ is proportional to Casimir's energy $E_C$.  Then, the total
energy $E$ should contain a sub-extensive term: $E = E_E + E_C/2$. The analysis
of the entropy with this total energy was performed for FRW models in \cite{ver}
and for Bianchi IX models in \cite{opq03}. For the cases considered here one 
can perform similar computations by allowing an additive constant term
in the entropy (\ref{tvar}), i.e. $S\rightarrow S + S_0$ with $S_0=$ const,
and then identifying $S_0$ with the Casimir energy. Then we obtain 
\be
S = \left[\frac{2\pi}{3} V^\omega \sqrt{2E_E E_C}\right]^{3/(2+3\omega)}\ .
\label{cardycos}
\ee
It is then possible to show that this expression for the total entropy 
still corresponds to an adiabatic process $\dot S =0$, if the dynamical
behavior of $E_E$ and $E_C$ is chosen correspondingly. So we see that
the sub-extensive character of Casimir's energy is not sufficient for
generating an entropy which would resemble the non-adiabatic character
of Cardy's entropy. 
%%%%%%%%%%%%%%%%%%%
Notice that the expression for the entropy of
the universe resembling the Cardy formula in terms of the different
kinds of energy, as given in Eq.(\ref{cardycos}), takes the special 
square--root form, originally postulated by Verlinde \cite{ver}, 
only in the case of a radiation dominated universe $(\omega = 1/3)$.
A similar result was obtained in \cite{youm02} in the limiting
FRW cosmologies.
%%%%%%%%%%%%%%%%%%% 

On the other hand, the adiabaticity condition for $S$ was obtained under
the assumption that the laws of relativistic thermodynamics are valid. 
According to the geometric
approach to thermodynamics described in Sections \ref{sec:class} and
\ref{sec:rel}, relativistic thermodynamics is obtained from its classical
version by assuming that the thermodynamic variables explicitly depend on the
coordinates used in the spacetime manifold that describes the gravitational
field. This procedure seems to lead to reasonable results in the
case of FRW and homogeneous (anisotropic) cosmologies, where the integrability 
conditions (\ref{ic2}) are trivially satisfied. However, in the
case of inhomogeneous cosmologies the resulting thermodynamic variables
are characterized by a very unphysical behavior \cite{kqs}. It is then
natural to ask whether the present version of 
relativistic thermodynamics is a definite and correct procedure. 
At least in the case of inhomogeneous fields a modification seems
to be necessary. 
The connection and curvature appear in the formalism only indirectly
through the spacetime coordinates. Perhaps one needs a generalized 
formalism which also takes into account the dynamics of curvature
and additional gravitational degrees of freedom.

Furthermore, 
a closer look at the first law of relativistic thermodynamics (\ref{gibbsdown1}) 
in the case of cosmological models considered here
 reveals that $U=E$  includes only the energy $\rho$ corresponding to the
perfect fluid and no ``gravitational energy" is taken into account. Of course,
this is a completely different problem since  
there is no general definition of energy for gravitational fields. 
Nevertheless, it seems
reasonable to demand that for a correct description of a thermodynamical system
one should include all kinds of energy present in the system.

%%%%%%%%%%%%%%%%%%%%%%%%%%%%%%%%
\section{Conclusions}
\label{sec:con}

In this work we used two different procedures to define the entropy of anisotropic
cosmological models in the case of Bianchi models of type I and V.
First, we presented a geometric approach to classical thermodynamics which allows us to
derive the relativistic generalization in a very simple manner. 
The relativistic version of the laws of thermodynamics 
was then used here to derive an expression for the entropy which is in accordance with 
an adiabatic cosmological expansion. This entropy shows a physical dynamical 
behavior an leads to the FRW case in the corresponding limit.
 
Secondly, we use the Cardy formula for the entropy of two-dimensional conformal
field theories and its generalization to any dimensions. It turns out that
for homogeneous cosmologies it is possible to choose the eigenvalue of the
Virasoro operator and the central charge in such a way that Cardy's entropy 
formula coincides with the Hamiltonian constraint for closed, plane, and 
open models. The resulting expression for Cardy's entropy, however, does
not correspond to an adiabatic expansion, even if it is considered 
as a sub-extensive thermodynamic variable. It is in this sense that we
conclude that it is not possible to interpret Cardy's entropy as the
entropy of a cosmological model. Nevertheless, we believe that this 
negative result is not sufficient to dismiss Cardy's entropy as
unphysical. The relativistic version of thermodynamics used to 
derive the adiabaticity condition leads to unphysical results, 
as soon as inhomogeneities are taken into account, so that
we cannot exclude the possibility of searching for a different
approach to relativistic thermodynamics.
 Also, this
version does not include all gravitational degrees of freedom
which intuitively are expected to affect the behavior of a 
thermodynamic system, 
%%%%%%%%%%%
especially its entropy. This would imply that there is no physical 
reason for demanding that Cardy's entropy should be comparable with
the thermodynamic entropy of a perfect fluid.

An additional point that one should examine critically is the
assumption of a perfect fluid as the source of the Bianchi anisotropic
models. A more realistic source must necessarily take into
account dissipative processes due to anisotropic expansions.
This would imply the analysis of more general fluids which
include viscosity terms, a task which is beyond the scope
of the present work.
%%%%%%%%%%

We conclude that it is necessary to perform a more detailed analysis 
in order to interpret Cardy's entropy as a thermodynamic variable which
is in accordance with the laws of relativistic thermodynamics.

\section{Acknowledgements}

It is a great pleasure
to dedicate this work to Octavio Obreg\'on on the ocassion of his 60-th birthday.
This work was supported in part by CONACyT grant 48601-F.

\end{document}